\documentclass[useAMS,usenatbib]{mn2e}

\usepackage[dvips]{graphicx} 

\title[M-I coupling at Jupiter-like exoplanets]{Magnetosphere-ionosphere coupling at Jupiter-like exoplanets with internal plasma sources: implications for detectability of auroral radio emissions}

\author[J.~D.~Nichols]{J.~D.~Nichols$^{1}$\thanks{E-mail:jdn@ion.le.ac.uk}\\
$^{1}$Department of Physics and Astronomy, University of Leicester, Leicester, LE1~7RH, UK}
\begin{document}

\date{Received 2011 February 10; in original form 2010 September 23}

\pagerange{\pageref{firstpage}--\pageref{lastpage}} \pubyear{2011}

\maketitle

\label{firstpage}

\begin{abstract}
	In this paper we provide the first consideration of magnetosphere-ionosphere coupling at Jupiter-like exoplanets with internal plasma sources such as volcanic moons.  We estimate the radio power emitted by such systems under the condition of near-rigid corotation throughout the closed magnetosphere, in order to examine the behaviour of the best candidates for detection with next generation radio telescopes.  We thus estimate for different stellar X-ray-UV (XUV) luminosity cases the orbital distances within which the ionospheric Pedersen conductance would be high enough to maintain near-rigid corotation, and we then consider the magnitudes of the large-scale magnetosphere-ionosphere currents flowing within the systems, and the resulting radio powers, at such distances.  We also examine the effects of two key system parameters, i.e.\ the planetary angular velocity and the plasma mass outflow rate from sources internal to the magnetosphere. In all XUV luminosity cases studied, a significant number of parameter combinations within an order of magnitude of the jovian values are capable of producing emissions observable beyond 1~pc, in most cases requiring exoplanets orbiting at distances between \ensuremath{\sim}1 and 50~AU, and for the higher XUV luminosity cases these observable distances can reach beyond \ensuremath{\sim}50~pc for massive, rapidly rotating planets.  The implication of these results is that the best candidates for detection of such internally-generated radio emissions are rapidly rotating Jupiter-like exoplanets orbiting stars with high XUV luminosity at orbital distances beyond \ensuremath{\sim}1~AU, and searching for such emissions may offer a new method of detection of more distant-orbiting exoplanets. 
\end{abstract}

\begin{keywords}
Planetary systems -- planets and satellites: aurorae, magnetic fields, detection.
\end{keywords}

\section{Introduction}
\label{sec:intro}

In recent years hundreds of exoplanets have been discovered, many of which (\ensuremath{\sim}21\%) have mass greater than or equal to that of Jupiter and orbital semi-major axes of $<0.1$~AU (where $\mathrm{1~AU\approx1.5\times 10^{11}~m}$), although a significant fraction (\ensuremath{\sim}61\%) of planets with Jupiter's mass or greater have been observed with semi-major axes $\ge1$~AU (see, e.g., the catalogue at exoplanet.eu).  The possibility of detection of the auroral radio emissions of `hot Jupiter'-like exoplanets close to their parent star has been considered by a number of authors \citep[e.g.][]{farrell99a, farrell04a, zarka01a,zarka07a,lazio04a, griessmeier04a, griessmeier05a, griessmeier07a, stevens05a, jardine08a, fares10a, reiners10a}. This interest has been sparked in part by the imminent commencement of high sensitivity radio observations by next generation radio telescopes such as the Low Frequency Array (LOFAR), which will have a detection threshold of 1 mJy (where $\mathrm{1\;Jansky = 10^{-26}\;W\;m^{-2}\;Hz^{-1}}$) \citep{farrell04a,griessmeier07a}. Such previous consideration of hot Jupiter-like exoplanets has assumed that the auroral radio emission would be caused by a star-planet interaction reminiscent of either the solar wind-Earth interaction or the Io-Jupiter interaction.  The former is mediated primarily via reconnection of the planetary and interplanetary magnetic fields at the dayside magnetopause \citep{dungey61}. This drives plasma flows within the magnetosphere that generate electric currents flowing between the magnetosphere and the resistive ionosphere, the upward magnetic field-aligned component of which, associated with downward-precipitating electrons, produces auroral and radio emissions.  The latter interaction is thought to be mainly associated with the generation of Alfv\'en waves in the vicinity of the moon Io, caused by its motion through the rapidly-rotating planetary magnetic field and plasma \citep{goertz80a,neubauer80a,crary97a}.  Consideration of such processes has led to the extrapolation of a `Radiometric Bode's Law' relating incident magnetic power to output radio power for the case of hot Jupiter-like exoplanets orbiting extremely close (at typically \ensuremath{\sim}10 stellar radii) of their parent stars.  It has been concluded that such interaction may generate emissions at or above the LOFAR detection threshold \citep{farrell04a, griessmeier07a}.  \\

\begin{figure}
 \noindent\includegraphics[width=84mm]{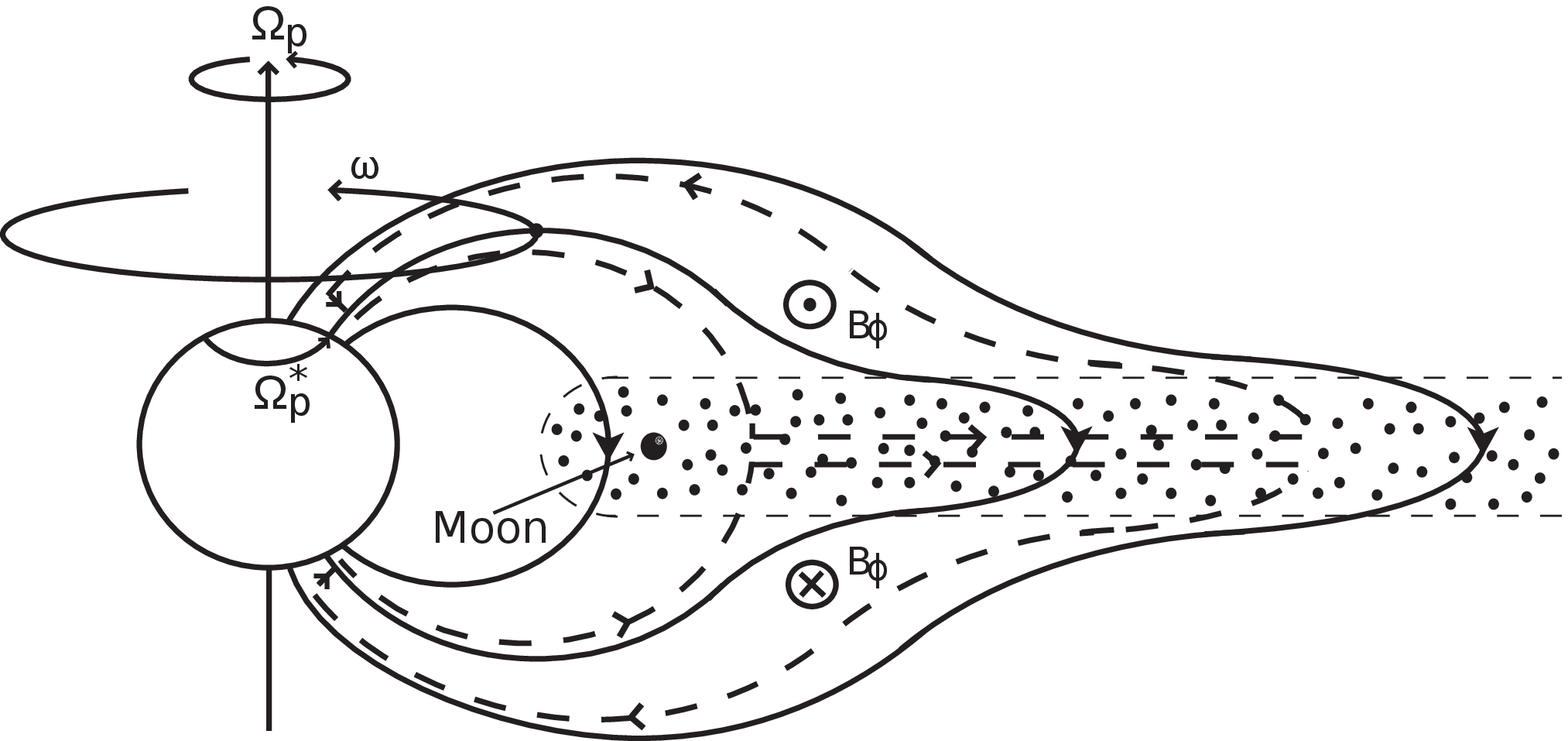}
\caption{
Sketch of a meridian cross section through a Jupiter-like exoplanet's inner and middle magnetosphere, showing the principal physical features involved. The arrowed solid lines indicate magnetic field lines, the arrowed dashed lines the magnetosphere-ionosphere coupling current system, and the dotted region the rotating disc of outflowing plasma.  After Cowley \& Bunce (2001).
}
\label{fig:miccs}
\end{figure}

However, despite the importance placed by previous authors on stellar wind-planet and Io-Jupiter type interactions, significant components of Jupiter's radio emissions, i.e.\ the b-KOM, HOM and non-Io-DAM emissions \citep{zarka98a}, are thought to be generated by the large-scale magnetosphere-ionosphere (M-I) coupling current system associated with the breakdown of corotation of iogenic plasma in Jupiter's middle magnetosphere, illustrated by Fig.~\ref{fig:miccs} \citep{hill79, hill01, pontius97, cowley01, nichols03, nichols04, nichols05}.  This process generates intense field-aligned electron beams which drive the brightest and most significant of Jupiter's UV auroral emission, i.e.\ the main auroral oval \citep{grodent03b, clarke04, nichols09b}, and, coupled with particle mirroring and the absorption of particles in the loss cone, excite the cyclotron maser instability in the high-latitude low-$\beta$ plasma, which gives rise to the above radio emissions. Observationally, the UV aurora and radio emissions of Jupiter and Saturn have been shown by a number of studies to be closely associated with one another \citep[e.g.][]{gurnett02,pryor05,kurth05a,clarke09a,lamy09a,nichols10a,nichols10b}.  Io orbits deep within Jupiter's magnetosphere at \ensuremath{\sim}5.9~\ensuremath{\mathrm{R_J}} (where \ensuremath{\mathrm{R_J}} represents the equatorial radius of Jupiter equal to 71,373~km), and its volcanoes liberate sulphur and oxygen atoms into a torus surrounding the moon's orbit at the rate of \ensuremath{\sim}1000$\;\mathrm{kg\;s^{-1}}$ \citep[e.g.][]{hill83, vasyliunas83, khurana93, bagenal97a, dols08a}.  These atoms are ionised by electron impact ionisation and thus become sensitive to the rotating planetary magnetic field, such that the newly-created plasma is picked up to corotation velocity. The picked-up plasma is centrifugally unstable and diffuses radially away from the planet, probably via flux-tube interchange motions \citep{siscoe81,pontius82,kivelson97,thorne97,bespalov06}, before being lost down the dusk flank of the magnetotail via the pinching off of plasmoids \citep[e.g.][]{vasyliunas83,woch02,vogt10a}.\\

As the plasma diffuses radially outward, its angular velocity drops (inversely with the square of the distance if no torques act) due to conservation of angular momentum, such that a radial gradient of angular velocity is set up in the equatorial plasma.  This angular velocity gradient, when mapped along the magnetic field to the ionosphere, causes an equatorward-directed (for Jupiter's magnetic field polarity) ionospheric Pedersen current to flow, the $\mathbf{J}\times\mathbf{B}$ force of which opposes the drag of the neutral atmosphere on the sub-rotating plasma. Angular momentum is transferred between the ionosphere and the equatorial plasma by the sweep-back of magnetic field lines into a lagging configuration, such that the ionospheric Pedersen current is balanced in the equatorial plane by an outward-directed (again, for Jupiter's magnetic field polarity) radial current, the $\mathbf{J}\times\mathbf{B}$ force of which tends to return the equatorial plasma back to corotation with the planet.  Current continuity between these two field-perpendicular currents is maintained via field-aligned (Birkeland) currents, the inner upward component of which is thought to generate Jupiter's main auroral oval emission \citep{cowley01,hill01,southwood01}.  The current system was studied in detail theoretically by \cite{nichols03}, who considered the effect of two poorly-constrained but important system parameters, the effective ionospheric Pedersen conductance \ensuremath{\Sigma_P^*}, and the plasma mass outflow rate \ensuremath{\dot{M}}, and they derived analytic approximations appropriate for small and large radial distances, the former of which will be instrumental in the present work.  \cite{nichols04} went on to examine the effect on the current system of modulation of the ionospheric Pedersen conductance due to auroral electron precipitation, and \cite{nichols05} and \cite{ray10a} have studied the effect of field-aligned voltages. \cite{cowley07} considered the effect on the current system of solar wind-induced expansions and contractions of Jupiter's magnetosphere, and recently the modulation of the current system by diurnal variation of the ionospheric Pedersen conductance caused by solar illumination has been considered by \cite{tao10a}.\\

\nocite{cowley01}
In this paper we consider the application of the model describing Jupiter's magnetosphere-ionosphere coupling current system to Jupiter-like exoplanets with internal plasma sources such as active moons.  The stability of satellites about exoplanets has been considered (see e.g. \cite{domingos06a} and references therein), since this is an important issue for the habitability of moons orbiting close-in `hot Jupiters'.  However, this is much less of a problem for  satellites about more distant-orbiting exoplanets, which will be shown to be important for the present study.  For example, \cite{barnes02a} could not place any mass limits on satellites orbiting planets beyond \ensuremath{\sim}0.6~AU. The vulcanism of Io is driven by tidal dissipation, which also acts to dampen the eccentricity of the moon's orbit, such that the small eccentricities of the orbits of the Galilean satellites of \ensuremath{\sim}0.001-0.01 are only raised due to Laplace resonance \citep{weiss04a}. The outer limit for stable orbits of a prograde moon was given by \cite{domingos06a} to be

\begin{equation}
	R_\mathrm{max}=a\left(\frac{M_p}{3M_\star}\right)^{1/3}\times0.4895(1-1.0305e_p-0.2738e_m)\;\;,
	\label{eq:rmax}
\end{equation}

\noindent where $a$ is the semi-major axis of the planet's orbit, $M_p$ is the mass of the planet, $M_\star$ is the mass of the parent star, $e_p$ is the eccentricity of the planet's orbit and $e_m$ is the eccentricity of the moon's orbit.  For values appropriate to Io (i.e. $a=5.2$~AU, $M_p=1.899\times 10^{27}$~kg, $M_\star=M_{\sun}=1.988\times 10 ^{30}$~kg, $e_p=0.048$, and $e_m=0.004$), $R_\mathrm{max}\simeq346~\ensuremath{\mathrm{R_J}}$, such that volcanic satellites orbiting jovian-like planets are expected to be stable (as is of course observed in the solar system, e.g. Io orbits at ~5.9 \ensuremath{\mathrm{R_J}}, well inside $R_\mathrm{max}$).  Indeed, for jovian mass planets orbiting beyond 1~AU, satellites with roughly Io's orbital distance will be typically stable, and only the planets with very eccentric orbits and high mass parent stars (e.g.\ $a=1$~AU, $M_{\star}=20M_{\sun}$, $e_p=0.75$, such that $R_\mathrm{max}\simeq 5.8~\ensuremath{\mathrm{R_J}}$) would be expected to be devoid of stable satellites at Io's orbital distance.\\

Both giant planets in our solar system which have been extensively studied by in-situ spacecraft, i.e.\ Jupiter and Saturn, possess moons which actively outgas into the near-planetary space.  At Jupiter this is Io, which emits material at the rate of \ensuremath{\sim}1000~\ensuremath{\mathrm{kg\;s^{-1}}} as discussed above, whilst at Saturn this is Enceladus, whose cryo-volcanoes emit water group ions at estimated rates ranging from a few to a few hundred~\ensuremath{\mathrm{kg\;s^{-1}}} \citep[e.g.][]{khurana07a,pontius06a}.  It is thus reasonable to assume that such active moons may be relatively prevalent amongst Jupiter-like exoplanets, and the resulting implications for detectability of auroral radio emissions is therefore considered here.  We begin from the assumption that Jupiter-like exoplanets are strongly illuminated by their parent star, such that the ionospheric Pedersen conductance is high enough to maintain near-rigid corotation throughout the magnetosphere, a condition which, as reviewed below, maximises the field-aligned current density for a given magnetosphere.  We then compute the currents, and thus the resulting radio power output, for varying configurations of Jupiter-like exoplanets.  Note that such strong illumination could be associated with close `hot Jupiters', but equally applies to more distant planets orbiting active stars. The parameters examined are the plasma mass outflow rate, planetary orbital distance, and rotation rate.  We show that, for planets with host stars more active than the Sun, only relatively modest modifications from the jovian system parameters are required to produce potentially-detectable configurations, and by doing so we open up the catalogue of potential candidates for detection by radio telescopes such as LOFAR to a class of planet previously overlooked. \\

\section{Theoretical background}
\label{sec:theory}

\begin{figure}
 \noindent\includegraphics[width=84mm]{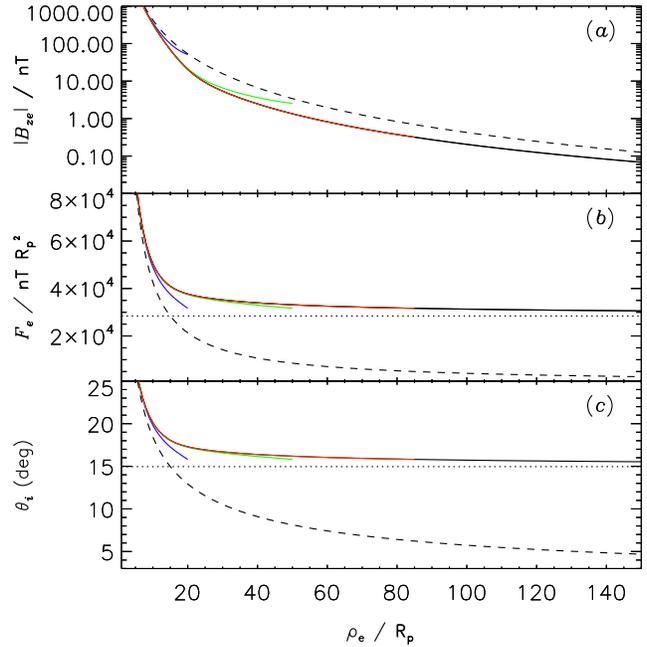}
\caption{
Plot showing (a) the magnitude of the north-south magnetic field strength threading the equatorial plane \ensuremath{|B_{ze}|} in nT, (b) the associated flux function $F_e$ in $\mathrm{nT~R_p^2}$, and (c) the ionospheric co-latitude $\theta_i$ to which the field lines threading the equatorial plane at radial distance \ensuremath{\rho_e} map. The red solid lines show the `baseline' model values given by Eqs.~\ref{eq:bze} and \ref{eq:fe}, while the black long-dashed lines show the planetary dipole values for comparison.  The other solid lines show the values modified by using Eqs.~\ref{eq:deltab} and \ref{eq:deltaf} for assumed magnetopause standoff distances \ensuremath{(R_{mp}/\mathrm{R_p}) \leq 85}, and by simply extending Eqs.~\ref{eq:bze} and \ref{eq:fe} for \ensuremath{(R_{mp}/\mathrm{R_p}) > 85}.  The colours blue, green, red, and black correspond to $(\ensuremath{R_{mp}}/\mathrm{R_p})$ values of 20, 50, 85, and $\geq 150$, respectively.  The horizontal dotted lines in panels (b) and (c) show the value of $F_\infty\approx 2.841\times 10^4$~nT~\ensuremath{\mathrm{R_p}^2}, and the corresponding limiting ionospheric co-latitude of the outer boundary of the model at \ensuremath{\sim 15\degr}.
}
\label{fig:bf}
\end{figure}

In this section we discuss the equations which describe the system, along with the approximations appropriate to the case of Jupiter-like exoplanets.  The analysis follows that which has been described previously by \cite{hill79}, \cite{pontius97}, \cite{cowley01}, and \cite{nichols03}, such that here we provide only a brief outline.  

\subsection{`Baseline' magnetic field model}

We start by outlining the magnetic field model, which acts as the basis for the current system.  It will become necessary to consider the variation of the current system with sub-solar magnetopause standoff distance \ensuremath{R_{mp}}, and here we thus describe the `baseline' model which is then modified by changing \ensuremath{R_{mp}}.  The magnetic field model is that which has been used previously for Jupiter by \cite{nichols04,nichols05}.  It is an axisymmetric poloidal field with the magnetic axis co-aligned with the spin axis, such that it can be described by the flux function $F(\rho,z)$, related to the magnetic field $\mathbf{B}$ by 
\begin{equation}
	\mathbf{B}=\left(\frac{1}{\rho}\right)\nabla F \times \hat{\varphi}\;\;,
	\label{eq:bf}
\end{equation}

\noindent where $\rho$ is the perpendicular distance from the magnetic axis, $z$ is the distance along the axis from the magnetic equator, and $\varphi$ is the azimuthal angle. Magnetic field lines are thus given by contours of $F$, such that mapping between the ionosphere (subscript i) and equator (subscript e) is simply achieved by writing $F_i=F_e$.  For a near-dipolar ionospheric field we have 
\begin{equation}
	F_i=B_{eq}\rho_i^2=B_{eq} R_p^2 \sin^2\theta_i\;\;,
	\label{eq:fi}
\end{equation}

\noindent where $\rho_i$ is the perpendicular distance from the magnetic axis in the ionosphere, $B_{eq}$ is the planet's equatorial dipole magnetic field strength (taken initially to be equal to Jupiter's value $B_J$~=~426,400~nT in conformity with the VIP4 planetary field model of \cite{connerney98}), $R_p$ is the planetary radius (taken to be equal to Jupiter's radius $R_J$), and $\theta_i$ is the magnetic co-latitude.  In this model, the magnitude of the north-south magnetic field threading the equatorial plane is given by
\begin{eqnarray}
	\lefteqn{B_{ze}^*(\rho_e)=}\nonumber \\ &-\left\{B_\circ^\prime\left(\frac{R_p}{\rho_e}\right)^3\exp\left[-\left(\frac{\rho_e}{\rho_{e\circ}}\right)^{5/2}\right]+B_\circ\left(\frac{R_p}{\rho_e}\right)^m\right\},
	\label{eq:bze}
\end{eqnarray}

\noindent where $B_\circ^\prime=3.335\times10^5$~nT, $\rho_{e\circ}=14.501$~\ensuremath{\mathrm{R_p}}, $B_\circ=5.4\times10^4$~nT, and $m=2.71$.  This form closely approximates the field model used by \cite{cowley01} and \cite{cowley02,cowley03a}, who employed the `Voyager-1/Pioneer-10' (`CAN') field model of \cite{connerney81} in the inner region and the Voyager-1 (`KK') model of \cite{khurana93} in the outer region.  This `baseline' field model is shown by the solid red line in Fig.~\ref{fig:bf}a.  The values of \ensuremath{|B_{ze}|} are always less than those of the pure planetary dipole, shown by the dashed line in Fig.~\ref{fig:bf}b, due to the radial stretching of the field lines associated with the current sheet.  The equatorial flux function is found by integration of Eq.~\ref{eq:bf} with the use of Eq.~\ref{eq:bze}, such that

\begin{eqnarray}
	\lefteqn{F_e^*(\rho_e)=}\nonumber\\
&	F_\infty+\frac{B_\circ^\prime R_p^3}{2.5\rho_{e\circ}}\Gamma\left[-\frac{2}{5},\left(\frac{\rho_e}{\rho_{e\circ}}\right)^{5/2}\right]+\frac{B_\circ}{(m-2)}\left(\frac{R_p}{\rho_e}\right)^{m-2},
	\label{eq:fe}
\end{eqnarray}

\noindent where $F_\infty\approx 2.841\times 10^4$~nT~\ensuremath{\mathrm{R_p}^2} is the value at infinity, and $\Gamma(a,z)=\int_z^\infty t^{a-1}e^{-t}\:dt$ is the incomplete gamma function.  This flux function is shown by the solid red line in Fig.~\ref{fig:bf}b, and is typically factors of \ensuremath{\sim}5 greater than the corresponding dipole flux function shown by the long-dashed line.

\subsection{Modification of the field structure by internal field strength and sub-solar magnetopause standoff distance}

We now consider how the above field model is modified by taking different values for the internal field strength and sub-solar magnetopause distance.  As summarised by \cite{griessmeier04a}, there are a number of models which estimate the planetary field strength given the planetary rotation rate, the radius of the dynamo region, and the mass density and conductivity within the dynamo region, although recently \cite{reiners10a} used an approach which was independent of the planetary rotation rate.  Here we will assume for simplicity jovian values for the planetary radius, core conductivity and mass density, such that the planetary field strength is solely dependent on the planetary rotation rate $\Omega_p$.  The planetary magnetic moment thus scales as $\mathcal{M}\propto {\Omega_P}^n$, where $n=1$ \citep{busse76a,sano93a}, $3/4$ \citep{mizutani92a}, or $1/2$ \citep{stevenson83a,mizutani92a}.  We thus take $n=3/4$ as representative of these choices, such that
\begin{equation}
	B_{ze}=\left(\frac{\Omega_P}{\Omega_J}\right)^{3/4}B_{ze}^*\;\;,
	\label{eq:bzeomega}
\end{equation}

\noindent and

\begin{equation}
	F_e=\left(\frac{\Omega_P}{\Omega_J}\right)^{3/4}F_e^*\;\;,
	\label{feomega}
\end{equation}

\noindent where $\Omega_J$ is Jupiter's planetary rotation period given by $\Omega_J=1.76\times10^{-4}\mathrm{\;rad\;s^{-1}}$, and $B_{ze}^*$ and $F_e^*$ are given by Eqs.~\ref{eq:bze} and \ref{eq:fe}, respectively. \\

Turning to the sub-solar magnetopause distance, this is expected to be dependent on the planet's orbital distance from the star via the stellar wind dynamic pressure \ensuremath{p_{sw}}, as discussed below, and on the internal field strength.  For the former, we employ the method of \cite{cowley07}, who studied the effect on the current system of changes to the magnetospheric size due to solar wind-induced compressions and expansions of Jupiter's magnetosphere. We note, however, that as well as considering the middle magnetosphere current sheet region described by the above field model, those authors also incorporated `outer magnetosphere' and open field line regions which are not included in this study since a significant flow shear at the open-closed field line boundary is not expected at planets with high ionospheric conductance, as considered here \citep{isbell84}.  Following \cite{cowley07}, we set the reference boundary of the `baseline' field model to be at 85~\ensuremath{\mathrm{R_p}}, as shown by the red lines in Fig.~\ref{fig:bf}, and modify the magnetic field model for smaller values of \ensuremath{R_{mp}}, by balancing the magnetic flux lost due to the closer outer boundary with the addition of the equivalent flux via a uniform southward field $\Delta B$ throughout the system, such that

\begin{equation}
	\pi {\ensuremath{R_{mp}}}^2 \Delta B=2\pi(F_e(\ensuremath{R_{mp}})-F_e(85\:\ensuremath{\mathrm{R_p}}))\;\;,
	\label{eq:deltab}
\end{equation}  

\noindent where $\Delta B$ thus represents the effect of the magnetopause currents as seen inside the magnetosphere.  The flux function $F$ is then equivalently modified by the addition of a term $\Delta F$ given by

\begin{equation}
	\Delta F=F_e(85\:\ensuremath{\mathrm{R_p}})-F_e(\ensuremath{R_{mp}})+\frac{{\Delta B(\ensuremath{R_{mp}}}^2-\rho_e^2)}{2}\;\;,
	\label{eq:deltaf}
\end{equation}

\noindent such that an identical amount of flux ($F_e(85\:\ensuremath{\mathrm{R_p}})\approx3.17\times 10^4$~nT~\ensuremath{\mathrm{R_p}^2}) is contained within the boundary of the model for any value of \ensuremath{R_{mp}} smaller than 85~\ensuremath{\mathrm{R_J}}, and thus the ionospheric location of the outer boundary is kept fixed at $\ensuremath{\sim}15.8\degr$.  For values of \ensuremath{R_{mp}} larger than 85~\ensuremath{\mathrm{R_p}}, we simply use $B_{ze}$ and $F_{e}$, since beyond \ensuremath{\sim}130 \ensuremath{\mathrm{R_p}} the magnitude of $\Delta B$ becomes larger than the more distant values of \ensuremath{|B_{ze}|}. Simply using $B_{ze}$ and $F_{e}$ to larger radial distances does result in a modest increase in the amount of flux contained within the model, thus modifying the ionospheric location of the outer boundary, but it is obvious from the difference between the lowest values of the red lines and the horizontal dotted lines in Figs.~\ref{fig:bf}b and c that this increase is very small, and the change in location of the open-closed field line boundary is less than $1\degr$, i.e.\ less than observed changes in the location of Jupiter's main auroral oval \citep{grodent08,nichols07, nichols09b}.  The magnetic field and flux values for these modified field structures are shown for magnetopause standoff distances of 20, 50, 85, and $\geq 150$ \ensuremath{\mathrm{R_p}} by the blue, green, red, and black lines in Fig.~\ref{fig:bf}, respectively.\\

We now discuss how the magnetopause standoff distance \ensuremath{R_{mp}} is expected to depend on planetary orbital distance \ensuremath{R_{orb}}.  The position of the magnetopause is dependent on the stellar wind dynamic pressure \ensuremath{p_{sw}=\varrho_{sw} \ensuremath{v_{sw}}^2}, where \ensuremath{v_{sw}} and $\varrho_{sw}$ are the stellar wind velocity and density, respectively.  These parameters are related to the mass loss rate of the star, which is measurable by detecting Ly-$\alpha$ absorption from the collision of the stellar wind with the surrounding interstellar medium \citep{wood05a}.  These authors showed that stars more active than the Sun (which may be most important for internally-generated exoplanetary radio emission, as discussed below) exhibit mass loss rates ranging between \ensuremath{\sim}0.01-100 times the solar value, and conjectured that although younger, more active stars are expected to exhibit greater mass loss rates, the high magnetic field strengths of the most active stars may in fact act to inhibit stellar wind outflow.  In view of these results, we employ solar wind velocity and density values as reasonably representative of the large observed range of stellar mass loss rates for active stars.  We thus assume that the stellar wind velocity \ensuremath{v_{sw}} is constant beyond 10 stellar radii at $\mathrm{450~km~s^{-1}}$, a value typical of those observed at both \ensuremath{\sim}1~AU \citep{hundhausen95a} and \ensuremath{\sim}5~AU \citep{nichols06}.  We then assume that the stellar wind density $\varrho_{sw}$ falls inversely with the square of the distance from the star (thus neglecting plasma sources in the interplanetary medium such as interstellar pick-up ions which become increasingly significant toward the outer regions of the heliosphere \citep[e.g.][]{intriligator96a,mccomas10a}).  The observed values of the solar wind density actually vary significantly over a solar rotation period due to the formation of corotating interaction regions in the interplanetary medium, and at Jupiter's orbit takes values between $\mathrm{0.01-1~cm^{-3}}$ \citep{nichols06}, with typical values being of order $\mathrm{0.1~cm^{-3}}$.  We thus employ the representative value of $\mathrm{0.1~cm^{-3}}$ at 5.2~AU, such that the stellar wind dynamic pressure varies with orbital distance as

\begin{equation}
	\ensuremath{p_{sw}}=0.91\left(\frac{1~\mathrm{AU}}{\ensuremath{R_{orb}}}\right)^2\;\mathrm{nPa}\;\;.
	\label{eq:psw}
\end{equation}

\noindent In the regime where the interplanetary magnetic field (IMF) strength is negligible with respect to the planetary field, stress balance between the dynamic pressure of the shocked stellar wind plasma in the magnetosheath adjacent to the magnetopause and the magnetic pressure of the compressed planetary vacuum dipole yields 

\begin{equation}
	\left(\frac{\ensuremath{R_{mp}}}{R_p}\right)=\left(\frac{k_m^2B_{eq}^2}{2\mu_\circ k_{sw} \ensuremath{p_{sw}}}\right)^{1/6}\;\;,
	\label{eq:rmp}
\end{equation}

\noindent where $k_{sw}\approx0.88$ for a high-Mach stellar wind \citep{spreiter70a} and $k_m$ represents the factor by which the magnetopause field is enhanced by magnetopause currents, given by $\ensuremath{\sim}2.44$ for a sub-solar boundary of realistic shape, i.e.\ lying between the values of 2 and 3 appropriate to planar and spherical boundaries, respectively \citep[e.g.][]{mead64a,alexeev05a}. Noting, however, that the precise values of $k_{sw}$ and $k_m$ are in practise unimportant due to the $1/6$ exponent in Eq.~\ref{eq:rmp}, this relation is approximated by

\begin{equation}
	\left(\frac{\ensuremath{R_{mp}}}{R_p}\right)\simeq \left(\frac{B_{eq}^2}{\mu_\circ \ensuremath{p_{sw}}}\right)^{1/6}\;\;,
	\label{eq:rmpdip}
\end{equation}

\noindent  similar relations to which have been used previously by various authors discussing exoplanetary magnetospheres \citep[e.g.][]{cuntz00a, zarka01a,griessmeier04a,griessmeier05a,griessmeier07a, jardine08a}.  However, it is well known that this formula underestimates the size of Jupiter's magnetosphere by a factor of \ensuremath{\sim}2, since Jupiter's magnetopause is inflated to \ensuremath{\sim}45-100~\ensuremath{\mathrm{R_J}}, with a mean of \ensuremath{\sim}75~\ensuremath{\mathrm{R_J}}, by the presence of internal plasma through the combined actions of thermal pressure and centrifugal force \citep{khurana04a}.  Therefore, \cite{huddleston98} presented an empirical relation for the variation of Jupiter's sub-solar magnetopause standoff distance \ensuremath{R_{mp}} with \ensuremath{p_{sw}}, given by\  
\begin{equation}
	\left(\frac{R_{mp}}{R_J}\right)=\frac{35.5}{(p_{sw}/\mathrm{nPa})^{0.22}}\;\;.
	\label{eq:hud}
\end{equation}

\begin{figure}
 \noindent\includegraphics[width=84mm]{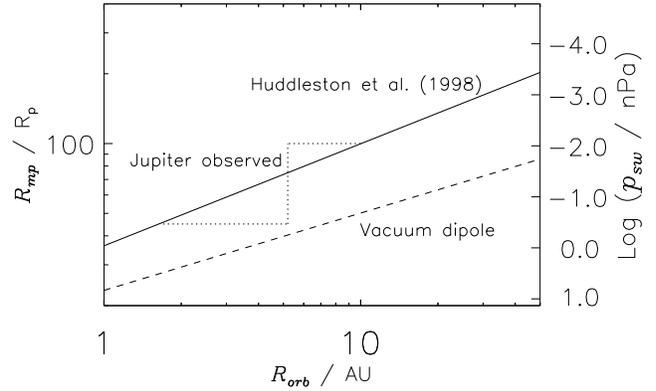}
\caption{
Plot showing (left axis) sub-solar magnetopause distance \ensuremath{R_{mp}} and (right axis) stellar wind dynamic pressure versus orbital radius \ensuremath{R_{orb}} computed using Eq.~\ref{eq:psw} and the Huddleston et al. (1998) empirical relation for Jupiter given by Eq.~\ref{eq:hud} (solid line) and the vacuum dipole relation given by Eq.~\ref{eq:rmpdip} (dashed line), both for jovian values of the planetary magnetic moment.  Also shown by the vertical dotted line at 5.2~AU is the range of magnetopause distances observed for Jupiter, thus the range over which the Huddleston et al. (1998) relation was defined, i.e.\ 45-100~\ensuremath{\mathrm{R_J}}, and the horizontal dotted lines show the orbital radii to which this range corresponds in the present study, i.e.\ \ensuremath{\sim}1.7-10~AU.
}
\label{fig:rmp}
\end{figure}
\nocite{huddleston98}
\noindent Exoplanets with active moons would similarly be expected to contain significant internal plasma, and here we thus employ a modification of Eq.~\ref{eq:hud}, taking into account the dependence on the internal field strength in Eq.~\ref{eq:rmpdip}, to estimate the values of \ensuremath{R_{mp}}, i.e.\

\begin{equation}
	\left(\frac{R_{mp}}{R_J}\right)=\frac{35.5}{(p_{sw}/\mathrm{nPa})^{0.22}}\left(\frac{B_{eq}}{B_J}\right)^{1/3}\;\;.
	\label{eq:hudmod}
\end{equation}

\noindent We show in Fig.~\ref{fig:rmp} the variation of sub-solar magnetopause distance \ensuremath{R_{mp}} and stellar wind dynamic pressure \ensuremath{p_{sw}} with orbital radius \ensuremath{R_{orb}} computed using Eq.~\ref{eq:psw} and the \cite{huddleston98} empirical relation for Jupiter given by Eq.~\ref{eq:hud} (solid line), along with the vacuum dipole relation given by Eq.~\ref{eq:rmpdip} (dashed line).  The vertical dotted line at 5.2~AU shows the range of sub-solar magnetopause distances observed for Jupiter, i.e.\ \ensuremath{\sim}45-100~\ensuremath{\mathrm{R_J}} corresponding to a solar wind dynamic pressure range of \ensuremath{\sim}0.01-0.32~nPa, hence representing the range over which the empirical \cite{huddleston98} relation is known to be valid.  The horizontal dotted lines then show the orbital radii to which this dynamic pressure range corresponds in the present study through Eq.~\ref{eq:psw}, i.e.\ \ensuremath{\sim}1.7-10~AU, although we acknowledge that intrinsic variations in the stellar wind dynamic pressure will actually induce a range in the sub-solar magnetopause standoff distances of an exoplanet at a given orbital distance, as observed at all solar system magnetospheres, an effect which we do not consider here. The validity of an empirical relation such as Eq.~\ref{eq:hud} beyond the observed values is, of course, not known.  However, the \cite{huddleston98} relation encapsulates in a relatively simple manner the response of a rotationally-driven, plasma-filled magnetosphere to changes in solar wind dynamic pressure, and produces results consistent with the 3D MHD simulations of Jupiter's magnetosphere by \cite{ogino98a} and, more crucially, with in situ observations of the boundary location.  On the other hand, the vacuum dipole relation given by Eq.~\ref{eq:rmpdip} represents an approximation at odds with a key property of the systems studied here, i.e. the radial distension of the field through the actions of the centrifugal force and internal plasma pressure, and does not provide results consistent with in situ observations of the jovian magnetosphere for any solar wind dynamic pressure value.  Thus, in the absence of any information about the size of extra-solar jovian-type magnetospheres, we use the expression validated by observations of Jupiter as a reasonable estimation, even though this sometimes involves extrapolation beyond the originally-observed range.

\subsection{Steady state plasma angular velocity and current system equations}
 
We next consider the angular velocity profiles arising from the application of Newton's second law to a radially-outward steady flow of plasma from the torus source.  As derived previously by \cite{hill79} and \cite{pontius97}, and discussed further by \cite{cowley02,cowley07} and \cite{nichols03,nichols04,nichols05}, the equatorial plasma angular velocity $\omega$ obeys the `Hill-Pontius' differential equation

\begin{equation}
	\frac{\rho_e}{2}\frac{\mathrm{d}}{\mathrm{d}\rho_e}\left(\frac{\omega}{\Omega_p}\right)+\left(\frac{\omega}{\Omega_p}\right)=\frac{4\pi \Sigma_P^*F_e|B_{ze}|}{\dot{M}}\left(1-\frac{\omega}{\Omega_p}\right)\;\;.
	\label{eq:hp}
\end{equation}

\noindent The left-hand side (LHS) of Eq.~\ref{eq:hp} represents the radial gradient of the plasma angular momentum per unit mass, whilst the right-hand side (RHS) is the ionospheric torque on the equatorial plasma per unit mass.  We also note that the effective Pedersen conductance \ensuremath{\Sigma_P^*} is reduced from the true value \ensuremath{\Sigma_P} by $\ensuremath{\Sigma_P^*}=(1-k)\ensuremath{\Sigma_P}$, where the parameter $k$ is related to the reduction of the angular velocity of the neutral atmosphere ($\Omega_p^*$) from rigid corotation ($\Omega_p$) via `slippage' \citep{huang89,millward05}, such that  $(\Omega_p-\Omega_P^*)=k(\Omega_p-\omega)$.  The value of $k$ is somewhat uncertain, so in common with previous works we take $k=0.5$.\\  

We now discuss the equations which describe the resulting magnetosphere-ionosphere coupling currents.  First, taking the ionospheric field to be vertical and equal to $2B_{eq}$ in strength (an approximation valid to within $\sim$5\% in our region of interest for planets with near-dipolar ionospheric fields \citep{nichols03}), the equatorward-directed height-integrated Pedersen current $i_P$ is given by

\begin{equation}
	i_P=2 \ensuremath{\Sigma_P^*}B_{eq}\Omega_p\rho_i\left(1-\frac{\omega}{\Omega_p}\right)\;\;.
	\label{eq:ip}
\end{equation}

\noindent Current continuity and the assumption of north-south symmetry then yields for the equatorial radial current integrated across the width of the current sheet $i_\rho$

\begin{equation}
	\rho_e i_\rho=2\rho_i i_P\;\;.
	\label{eq:irhoip}
\end{equation}

\noindent From Eqs.~\ref{eq:ip}, ~\ref{eq:irhoip} and \ref{eq:fi}, and recalling that $F_i=F_e$ we have 

\begin{equation}
	i_\rho=\frac{4 \ensuremath{\Sigma_P^*}F_e\Omega_p}{\rho_e}\left(1-\frac{\omega}{\Omega_p}\right)\;\;,
	\label{eq:irho}
\end{equation}

\noindent such that the total radial current integrated in azimuth $I_\rho$ is

\begin{equation}
	I_\rho=2\pi\rho_ei_\rho=8\pi \ensuremath{\Sigma_P^*}\Omega_p F_e \left(1-\frac{\omega}{\Omega_p}\right)\;\;,
	\label{eq:totip}
\end{equation}

\noindent equal to twice the azimuth-integrated Pedersen current $I_P = 2\pi\rho_ii_P$ flowing in each hemisphere.  The field-aligned current density, e.g., northward out of the northern surface of the equatorial current sheet is then given, through current continuity, by the divergence of the radial current, such that

\begin{equation}
	j_{ze}=-\frac{1}{2}\,\nabla{}\cdot{i_\rho}=-\frac{1}{2\rho_e}\frac{\mathrm{d}}{\mathrm{d}\rho_e}({\rho_ei_\rho})\;\;,
	\label{eq:jz}
\end{equation}

\noindent where the factor of a half arises from the assumption of hemispheric symmetry, i.e.\ an equal and opposite current flows out of the southern face of the current sheet.  Noting that the quantity $(j_\|/B)$ is constant along a field line, such that $(j_{\|i}/(2\ensuremath{B_{eq}})) = (j_{ze}/ \ensuremath{B_{ze}})$, we find the following expression for the field-aligned current density at the top of the ionosphere

\begin{equation}
	\ensuremath{j_{\|i}}=\frac{B_{eq}}{2\pi\rho_e|B_{ze}|}\frac{\mathrm{d}I_\rho}{\mathrm{d}\rho_e}\;\;.
	\label{eq:jpari}
\end{equation}

\noindent With these currents in mind, we now consider the approximations appropriate to strongly-irradiated Jupiter-like exoplanets.

\subsection{High-conductance approximation and conductance estimations}

\begin{figure}
 \noindent\includegraphics[width=84mm]{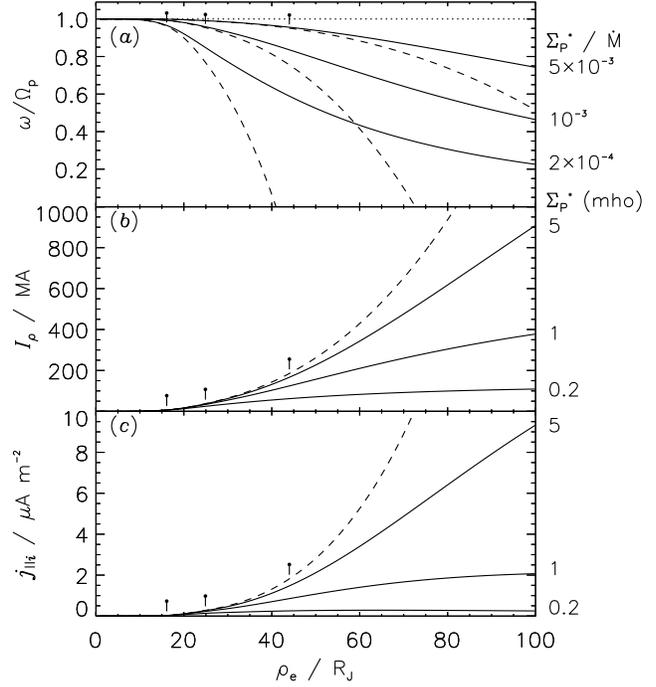}
\caption{
Plot showing profiles of the current system parameters using the full solution of Eq.~\ref{eq:hp} (solid lines) and the high-conductance approximation given by Eq.~\ref{eq:omegas} (dashed lines).  Parameters shown are (a) the equatorial plasma angular velocity, (b) the azimuthally-integrated equatorial radial current, and (c) the field-aligned current density. Three results are shown for $(\ensuremath{\Sigma_P^*}/\ensuremath{\dot{M}})=\mathrm{5\times10^{-3},10^{-3},~and~2\times10^{-4}~ mho~s~kg^{-1}}$, with \ensuremath{\dot{M}} set to 1000~\ensuremath{\mathrm{kg\;s^{-1}}} in panels b and c.  Note only one dashed line is shown in panels b and c since the high-conductance approximations (given by Eqs.~\ref{eq:totirhos} and \ref{eq:jparis} for the radial and field-aligned currents, respectively) are independent of \ensuremath{\Sigma_P^*}.  The tick marks show the limit of validity of the approximation as suggested by Nichols \% Cowley (2003).  
}
\label{fig:omega}
\end{figure}

Equation~\ref{eq:hp} is a first order, linear, inhomogeneous ordinary differential equation, which can be solved analytically for power law magnetic field models, such as a dipole field \citep{hill79,hill01} or the `KK' field model valid throughout the majority of Jupiter's middle magnetosphere \citep{nichols03}, and numerically for other field models.  The solutions have the property that they are dependent on the quotient $(\ensuremath{\Sigma_P^*}/\ensuremath{\dot{M}})$, as shown in Fig.~\ref{fig:omega}a for roughly jovian values of the ionospheric Pedersen conductance \ensuremath{\Sigma_P^*} and equatorial plasma mass outflow rate \ensuremath{\dot{M}}.  Specifically, the typical scale over which the plasma falls from rigid corotation is called the `Hill distance' $\rho_H$ after T.~W.~Hill, who first derived this scale length for a dipole magnetic field (not to be confused with the radius of the Hill sphere within which a body's gravitational field is dominant). The scale length was modified by \cite{nichols03} for the case of a magnetic field that varies as a power law with arbitrary exponent $m$, and for which the field lines map from the equatorial plane to a narrow band in the ionosphere.  This is appropriate for the stretched current sheet magnetic field of Jupiter's middle magnetosphere, as can be appreciated from the ionospheric mapping shown in Fig.~\ref{fig:bf}c, in which between 20-150~\ensuremath{\mathrm{R_J}} the current sheet field models indicated by the solid lines map between \ensuremath{\sim}15-17\degr, whereas the dipole field indicated by the dashed line maps to a much broader region between \ensuremath{\sim}5-13\degr.  The current sheet scale length is given by

\nocite{nichols03}
\begin{equation}
	\left(\frac{\rho_{H}}{R_p}\right) = \left(\frac{2\pi\Sigma_P^*B_\circ F_\circ}{\dot{M}}\right)^{1/m}\;\;,
	\label{eq:rhoh}
\end{equation}

\noindent where $F_\circ$ is the value of the flux function at the narrow ionospheric band, taken here to be equal to $F_\infty$, shown by the dotted horizontal line in Fig.~\ref{fig:bf}.  Thus, for a given equatorial distance, the plasma angular velocity increases for increasing values of $(\ensuremath{\Sigma_P^*}/\ensuremath{\dot{M}})$.  \cite{nichols03} therefore derived analytic approximations to the solution appropriate to both the inner region, where the plasma near-rigidly corotates, and the outer region, where the plasma can be considered to be either free of ionospheric torque or stagnant.  The inner region approximation for the equatorial plasma angular velocity, which \cite{nichols03} suggested is valid out to a distance $\rho_{S\:\mathrm{lim}}$ of

\begin{equation}
	 \rho_{S\:\mathrm{lim}}=0.516\left(\frac{\rho_H}{R_P}\right)\;\;,
	\label{eq:slim}
\end{equation}

\noindent is the leading term in the series solution of Eq.~\ref{eq:hp}, taking $(\ensuremath{\dot{M}}/\Sigma_P^*)$ as the formal expansion parameter and $(\omega/\Omega_p)=1$ at $\rho_e=0$ (see \cite{nichols03} for further details), such that 

\begin{equation}
	\left(\frac{\omega}{\Omega_p}\right)_S=1-\frac{1}{2}\left(\frac{\rho_e}{\rho_H}\right)^m\;\;,
	\label{eq:omegas}
\end{equation}

\noindent as shown by the dashed lines in Fig.~\ref{fig:omega}a.  Substitution of Eq.~\ref{eq:omegas} into Eqs.~\ref{eq:totip} and \ref{eq:jpari} thus yields

\begin{equation}
	I_{\rho\:S}=2I_{P\:S}=\frac{2\dot{M}\Omega_p}{|B_{ze}|}\;\;,
	\label{eq:totirhos}
\end{equation} 

\noindent and

\begin{equation}
	\ensuremath{j_{\|i\:S}}=-\frac{\dot{M}\Omega_p B_{eq}}{\pi \rho_e |B_{ze}|^3}\frac{\mathrm{d}|B_{ze}|}{\mathrm{d}\rho_e}\;\;,
	\label{eq:jparis}
\end{equation}

\noindent for the total radial current and ionospheric field-aligned current density, respectively.  These limiting currents, which are dependent only on \ensuremath{\dot{M}} and not \ensuremath{\Sigma_P^*}, are those required to maintain near-rigid corotation throughout the magnetosphere, and can thus be viewed alternatively as a high-\ensuremath{\Sigma_P^*} approximation for a given value of \ensuremath{\dot{M}}.  \\

The currents $I_{\rho}$ and \ensuremath{j_{\|i}} derived from the angular velocity profiles shown in Fig.~\ref{fig:omega}a are shown by the solid lines in Figs.~\ref{fig:omega}b and c, respectively, for roughly jovian field and current system parameters, i.e.\ \ensuremath{\dot{M}}=1000~\ensuremath{\mathrm{kg\;s^{-1}}} and \ensuremath{\Sigma_P^*}=0.2, 1, and 5 mho.  It is apparent that the current magnitudes increase with increasing \ensuremath{\Sigma_P^*}, asymptoting toward the limiting currents shown by the dashed lines.  Thus, the high-conductance approximation represents the condition under which maximum currents, and therefore maximum output radio power, are generated.  The minimum conductance required such that the high-conductance approximation is valid throughout the magnetosphere is found by setting $\ensuremath{R_{mp}}=\rho_{S\:\mathrm{lim}}$ from Eq.~\ref{eq:slim}, such that

\begin{equation}
	\left(\frac{\ensuremath{R_{mp}}}{\ensuremath{R_p}}\right)=0.516 \left(\frac{\rho_H}{\ensuremath{R_p}}\right)\;\;,
\end{equation}

\noindent which, upon substitution of Eq.~\ref{eq:rhoh} and rearranging for \ensuremath{\Sigma_P^*}, yields

\begin{equation}
	\Sigma_{P\;lim}^*\approx0.956\frac{\dot{M}}{B_\circ F_\circ}\left(\frac{R_{mp}}{R_p}\right)^m\;\mathrm{mho}\;\;.
	\label{eq:spslim}
\end{equation}

\noindent  Since we are interested in current systems powerful enough to be detected across interstellar distances, as possibly possessed by Jupiter-like exoplanets strongly irradiated by their parent stars, such a high-conductance is clearly appropriate and will be used as the basis for the results which follow.\\

Before considering the precipitating electron energy flux and associated radio power output, it is appropriate here to discuss the region of validity of the high-conductance approximation by estimating how the ionospheric Pedersen conductance varies with planetary orbital distance \ensuremath{R_{orb}}, and the type of star. We use a model for the Pedersen conductance which comprises two components, i.e.\ conductance produced by solar illumination, and auroral particle precipitation.  For the former, in the absence of more complex ionospheric models for Jupiter-like exoplanets we assume that the Pedersen conductance is, like the Earth's, linearly proportional to the ionospheric electron number density, a quantity proportional to the square root of the electron production rate, which is in turn linearly proportional to the solar extreme ultraviolet flux \citep{robinson84a,luhman95a}, although we note that actually the Earth's upper atmosphere is ionised by solar radiation from X-ray to FUV wavelengths (a range denoted by the term `XUV'). Thus, we have for the Pedersen conductance generated by stellar illumination 

\begin{equation}
	\Sigma_{P\:XUV}^*=\left(\frac{L_{XUV\:\star}}{L_{XUV\:\sun}}\right)^{1/2}\frac{2.6}{\ensuremath{R_{orb}}}\;\mathrm{mho}\;\;,
	\label{eq:speuv}
\end{equation} 

\noindent where $L_{XUV\:\star}$ and $L_{XUV\:\sun}$ are the XUV luminosities of the exoplanet's parent star and the Sun, respectively, and where the constant of proportionality is set such that $\Sigma_{P\:XUV}^*~=~0.5$~mho at 5.2~AU for solar XUV luminosity, consistent with estimates for Jupiter \citep[e.g.][]{achilleos98}.  The Pedersen conductance is therefore expected to be higher for stars with higher XUV luminosities, such as more active stars.  For example, the \ensuremath{\sim}170~nm flux from F2V-type stars is two orders of magnitude larger than solar-type G2V stars \citep{kasting96a}, since G2V type stars are sub-luminous in the FUV by a factor of ten or more due to absorption in this waveband \citep{lean97a}.   Luminosity at X-ray-EUV wavelengths is mainly non-thermal coronal emission, and surveys indicate that most coronal sources in the EUV are active stars with coronal X-ray and EUV luminosities several orders of magnitude greater than that of the Sun \citep{barstow03a}.  Stellar age is an important factor which determines stellar X-ray-EUV luminosity \citep{ribas05a}, with younger main sequence stars typically exhibiting higher X-ray-EUV emissions.  In the results which follow, we will thus consider stellar XUV luminosities from 1 to $10^3 L_{XUV\:\sun}$, a range consistent with values observed by \cite{ribas05a} for solar-type stars, and thus could be conservative with respect to other spectral types.  \\

As well as stellar illumination, Pedersen conductance is expected to be generated by auroral electron precipitation.  Computations by \cite{millward02} indicated that up to \ensuremath{\sim}8~mho is generated by auroral precipitation at Jupiter, and \cite{nichols04} used these results to compute the expected variation of conductance with field-aligned current density assuming a representative primary and secondary precipitating electron spectrum.  In their model the conductance plateaus at \ensuremath{\sim}3~mho at \ensuremath{\sim}0.5~\ensuremath{\umu \mathrm{A\;m^{-2}}}, since at higher currents the primary electrons precipitate lower than the Pedersen conducting layer and the conductance is generated by the lower energy secondaries.  A full analysis of the effect of precipitation induced conductance requires that Eq.~\ref{eq:hp} and Eq.~\ref{eq:jpari} are solved simultaneously, as in the computations of \cite{nichols04}, but here we simply assert that \ensuremath{j_{\|i}} is everywhere greater than 0.5~\ensuremath{\umu \mathrm{A\;m^{-2}}}, as is generally the case for the powerful exoplanetary systems we consider below, such that we assume a constant auroral contribution to the conductance of 3~mho, such that 

\begin{equation}
	\Sigma_{P}^*=\left(\frac{L_{XUV\:\star}}{L_{XUV\:\sun}}\right)^{1/2}\frac{2.6}{\ensuremath{R_{orb}}}+3\;\mathrm{mho}\;\;.
	\label{eq:sps}
\end{equation}

\noindent The maximum orbital distance \ensuremath{R_{orb}^*} for which the high-conductance approximation is valid is thus obtained simply by setting \ensuremath{\Sigma_{P\;lim}^*} given by Eq.~\ref{eq:spslim} equal to \ensuremath{\Sigma_P^*} given by Eq.~\ref{eq:sps} and solving for \ensuremath{R_{orb}}, numerically in practice.  This then limits the maximum power output for a given system, since at larger distances the high-conductance approximation is not valid and the magnetosphere will not rigidly corotate, and, as will be seen below, at smaller distances the power required to maintain the more compressed magnetosphere rigidly rotating is reduced.

\begin{figure}
\noindent\includegraphics[width=84mm]{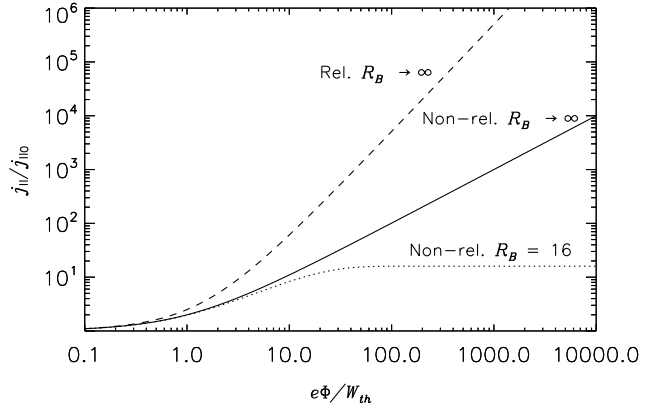}
\caption{
Plot showing the field aligned current \ensuremath{j_{\|i}} normalised to the limiting current \ensuremath{j_{\|i\circ}} produced by field-aligned voltage $\Phi$ normalised to the unaccelerated population thermal energy $W_{th}$, for the three current-voltage relations given by Eqs.~\ref{eq:jfk} for $R_B=16$, appropriate for an accelerator height of \ensuremath{\sim}2.5~\ensuremath{\mathrm{R_J}} (dashed line), \ref{eq:jkt} (solid line), and \ref{eq:jrk} (dotted line).  ``Rel.'' means ``relativistic'', ``Non-rel.'' means ``non-relativistic'', and $R_B$ is the ratio of the ionospheric magnetic field strength to the field strength at the top of the voltage drop.
}
\label{fig:jp}
\end{figure}

\subsection{Parallel voltage, precipitating electron energy flux, and radio power output}

We now briefly discuss the precipitating electron energy flux and associated radio power output that results from the field-aligned current given by Eq.~\ref{eq:jparis}.  The maximum field-aligned current that can be carried by unaccelerated precipitating magnetospheric electrons is 

\begin{equation}
	\ensuremath{j_{\|i\circ}}=eN \left(\frac{W_{th}}{2\pi m_e}\right)^{1/2}\;\;,
\end{equation}

\noindent with a corresponding precipitating energy flux of

\begin{equation}
	\ensuremath{E_{f\circ}}=2NW_{th}\left(\frac{W_{th}}{2\pi m_e}\right)^{1/2}\;\;,
\end{equation}

\noindent where $e$ is the elementary charge, $N$ is the magnetospheric electron number density, $W_{th}$ is the electron thermal energy, and $m_e$ is the electron mass.  In the absence of detailed models or empirical values for exoplanetary magnetospheric plasmas we take, in common with previous works, jovian values for the plasma parameters outside the current sheet, i.e.\ $N=0.01\;\mathrm{cm^{-3}}$ and $W_{th}=2.5\;\mathrm{keV}$ \citep{scudder81}, such that \ensuremath{j_{\|i}\approx0.013\;\mathrm{\umu A\;m^{-2}}}.  A significant field-aligned voltage is thus required to drive field-aligned currents greater than this value.  The current-voltage relation was first derived by \cite{knight73} and is given, for an arbitrary value of the ratio $R_B$ of the ionospheric magnetic field strength to the field strength at the top of the voltage drop, by

\begin{equation}
	\left(\frac{\ensuremath{j_{\|i}}}{\ensuremath{j_{\|i\circ}}}\right)=
	1 + (R_B - 1)\exp \left(-\frac{e\Phi}{W_{th}(R_B-1)}\right)\;\;,
	\label{eq:jfk}
\end{equation}

\noindent where $\Phi$ is the field-aligned voltage.  In the limit that the field-aligned voltage region is located at high altitude (typically $> 4\;\ensuremath{\mathrm{R_J}}$ up the field lines at Jupiter), the ratio $R_B$ becomes large (e.g.\ $R_B=216$ at $6\;\ensuremath{\mathrm{R_J}}$) such that Eq.~\ref{eq:jfk} reduces to

\begin{equation}
	\left(\frac{\ensuremath{j_{\|i}}}{\ensuremath{j_{\|i\circ}}}\right)=
	1 + \left(\frac{e\Phi}{W_{th}}\right)\;\;,
	\label{eq:jkt}
\end{equation}

\noindent the relation used by \cite{cowley01,cowley02,cowley07,nichols03,nichols04,nichols05}.  On the other hand, \cite{ray10a} suggested that the accelerator would be fixed at low altitude (2-3~\ensuremath{\mathrm{R_J}}), such that they employed Eq.~\ref{eq:jfk} with $R_B=16$.  However, the \cite{knight73} relations given by Eqs.~\ref{eq:jfk} and \ref{eq:jkt} assume the accelerating voltage is small such that the precipitating particles are accelerated to non-relativistic speeds, an assumption which is not valid when the field-aligned voltage approaches or exceeds the electron rest energy (\ensuremath{\sim}511~keV), as would be the case for powerful exoplanetary systems, as shown below.  This restriction was removed by \cite{cowley06b}, who showed that for a relativistic accelerated population the current-voltage relation becomes, in the high $R_B$ limit,

\begin{equation}
	\left(\frac{\ensuremath{j_{\|i}}}{\ensuremath{j_{\|i\circ}}}\right)=
	1 + \left(\frac{e\Phi}{W_{th}}\right)+
	\frac{\left(\frac{e\Phi}{W_{th}}\right)^2}{2\left[\left(\frac{m_ec^2}{W_{th}}\right)+1\right]}\;\;,
	\label{eq:jrk}
\end{equation}

\noindent where $c$ is the speed of light.  The current-voltage relations given by Eqs.~\ref{eq:jfk}, \ref{eq:jkt}, and \ref{eq:jrk} are shown in Fig.~\ref{fig:jp}, in which it is apparent that the relativistic and fixed-height accelerator cases produce competing deviations from the simple relation given by Eq.~\ref{eq:jkt}.  In light of the uncertainty surrounding the exact plasma parameters at Jupiter-like exoplanets, we thus employ the limiting \cite{knight73} relation given by Eq.~\ref{eq:jkt} and shown by the solid line in Fig.~\ref{fig:jp}, as providing a reasonable representation of the available choices for the current-voltage relation, and we simply note that this relation has been successfully used previously to compute auroral brightness and precipitating electron energy values for Jupiter that are consistent with observations \citep[e.g.][]{cowley01}.\\

The precipitating energy flux associated with the electrons accelerated through voltage $\Phi$ as given by Eq.~\ref{eq:jkt} was shown by \cite{lundin78a} to be

\begin{equation}
	\ensuremath{E_f}=
	\frac{\ensuremath{E_{f\circ}}}{2}\left[\left(\frac{\ensuremath{j_{\|i}}}{\ensuremath{j_{\|i\circ}}}\right)^2+1\right]\simeq
		\frac{\ensuremath{E_{f\circ}}}{2}\left(\frac{\ensuremath{j_{\|i}}}{\ensuremath{j_{\|i\circ}}}\right)^2\;\;,
		\label{eq:ef}
\end{equation}

\noindent where the approximation on the RHS is valid where $\ensuremath{j_{\|i}}~\gg~\ensuremath{j_{\|i\circ}}$, a condition that is valid essentially everywhere in the models we investigate.   The total precipitating power for each hemisphere $P_e$ is thus obtained by integration of \ensuremath{E_f} over the hemisphere, such that

\begin{equation}
	P_e=\int_0^{90}2\pi {R_p}^2\sin\theta_i \;E_f \;d\,\theta_i\;\;.
	\label{eq:pe}
\end{equation}

\noindent  Although the exact details of the mechanism by which radio frequency waves are generated from unstable auroral electron distributions remain to be determined, the most likely candidate is the electron cyclotron maser  instability \citep[e.g.][]{wu79a}, which has a generation efficiency of \ensuremath{\sim}1\% \citep{zarka98a}, such that we take the total emitted radio power to be given by

\begin{equation}
	P_r=\frac{P_e}{100}\;\;.
	\label{eq:pr}
\end{equation}

\begin{figure}
\noindent\includegraphics[width=84mm]{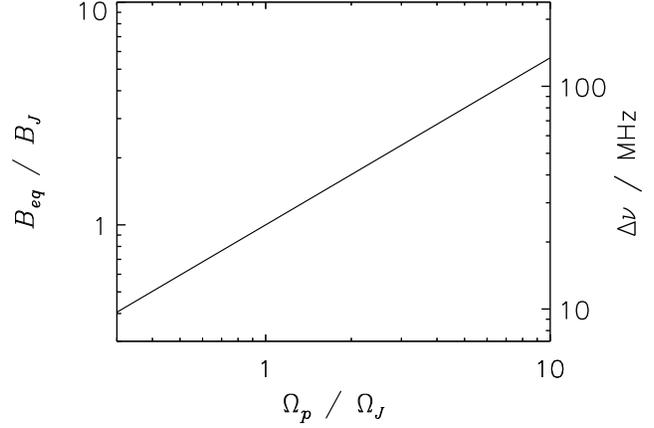}
\caption{
Plot showing (on the left axis) the equatorial surface magnetic field strength \ensuremath{B_{eq}} normalised to Jupiter's value $B_J$, and (on the right axis) the estimated radio emission bandwidth versus planetary angular velocity over the range $0.3<\ensuremath{\Omega_p}/\ensuremath{\Omega_J}<10$, computed from Eqs.~\ref{eq:bzeomega} and \ref{eq:delnu}, respectively. 
}
\label{fig:obf}
\end{figure}

\noindent We finally assume that the radio emission is beamed into solid angles of 1.6~sr, in conformity with observations of the jovian HOM and DAM emissions \citep{zarka04a}.  The effect of beaming is to increase the flux at a given distance but decrease the chance of detection, and here we assume that the beam from only one hemisphere is observable at any one time.  Assuming such beamed emission over the bandwidth $\Delta\nu$, the spectral flux density at a given distance $s$ from the source is

\begin{equation}
	F_r=\frac{P_r}{1.6 s^2 \Delta \nu}\;\;,
	\label{eq:fr}
\end{equation}

\noindent where the bandwidth $\Delta\nu$ is taken to be equal to the electron cyclotron frequency in the polar ionosphere, i.e.\ 

\begin{equation}
	\Delta\nu=\frac{eB_{eq}}{\pi m_e}\;\;,
	\label{eq:delnu}
\end{equation}

\noindent again assuming a uniform polar ionospheric field strength equal to $2B_{eq}$. We show in Fig.~\ref{fig:obf} the field strengths and resulting estimated radio bandwidths for the range of planetary angular velocities considered in this paper (see below).

\section{Results}
\label{sec:results}

\begin{figure}
\noindent\includegraphics[width=84mm]{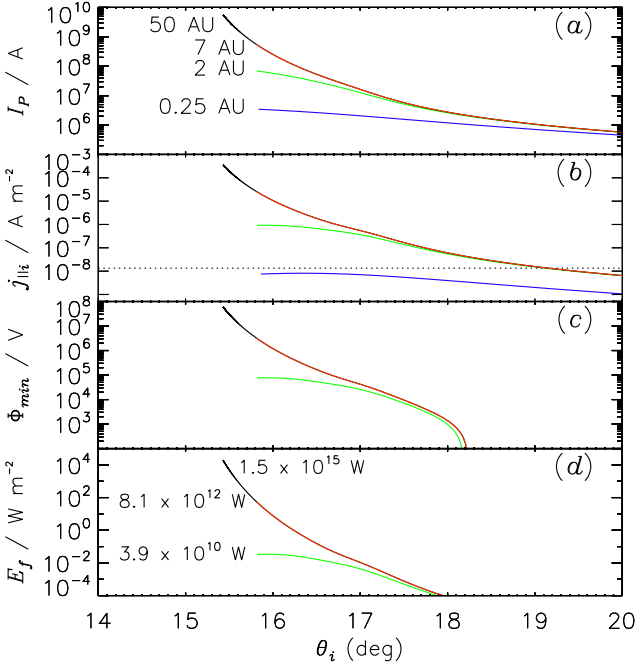}
\caption{
Plot showing the current system parameters versus ionospheric co-latitude, for magnetopause standoff distances of \ensuremath{R_{mp}=20,50,85\mathrm{, and~}200~\ensuremath{\mathrm{R_p}}} (blue, green, red, and black lines respectively), and with $\Omega_p=\Omega_J$ and $\dot{M}=\dot{M}_J$.  The labels in panel (a) correspond to the values of \ensuremath{R_{orb}} which result in magnetospheres of these sizes, according to Eqs.~\ref{eq:psw} and \ref{eq:hud}. Parameters shown are (a) the azimuthally-integrated ionospheric Pedersen current $I_P$ in A as given by Eq.~\ref{eq:totirhos}, (b) the field-aligned current density at the top of the ionosphere \ensuremath{j_{\|i}} in \ensuremath{\mathrm{A\;m^{-2}}} given by Eq.~\ref{eq:jparis}, (c) the minimum field-aligned voltage $\Phi_{min}$ in V required to drive the current in panel (b) as given by Eq.~\ref{eq:jkt}, and (d) the precipitating electron energy flux $E_f$ in \ensuremath{\mathrm{W\;m^{-2}}} as given by Eq.~\ref{eq:ef}. The labels in panel (d) show the radio power obtained by integrating the energy fluxes using Eqs.~\ref{eq:pe} and \ref{eq:pr}. Note there are no blue lines in panels (c) and (d) since for this case the current density in panel (b) is everywhere below the minimum current \ensuremath{j_{\|i\circ}} for which field-aligned voltages are required (shown by the horizontal dotted line in panel (b)).  
}
\label{fig:is}
\end{figure}

We now consider the application of the analysis presented in Sec.~\ref{sec:theory} to various configurations of exoplanets.  We first show in Fig.~\ref{fig:is} the effect on the current system of taking different values of \ensuremath{R_{mp}}, whilst maintaining jovian values for the planetary angular velocity $\Omega_p=\Omega_J$ (and thus $B_{eq}=B_J$) and mass plasma outflow rate $\dot{M}=\dot{M}_J=1000~\mathrm{kg~s^{-1}}$, thus simulating transplanting the jovian system to different orbital distances, assuming that the XUV flux is at all distances sufficiently high such that high-conductance approximation holds everywhere.  It is apparent from Fig.~\ref{fig:is} that for values of \ensuremath{R_{mp}} less than the reference value of 85~\ensuremath{\mathrm{R_J}} the current amplitudes are, as expected from e.g. Eq.~\ref{eq:totirhos}, reduced owing to the elevated magnetic field strength \ensuremath{|B_{ze}|}, which we recall is increased via Eq.~\ref{eq:deltab} for a smaller magnetosphere due to flux conservation.  For these model parameters, very compressed magnetospheres with \ensuremath{R_{mp}} smaller than \ensuremath{\sim}20~\ensuremath{\mathrm{R_J}}, are such that the field-aligned current density is everywhere less than \ensuremath{j_{\|i\circ}}, hence no field-aligned voltages are required and there is no bright discrete auroral emission.  Hence, even if we neglect tidal-locking which would in reality act to prevent the fast rotation of such planets (discussed further below), very close-orbiting ($\ensuremath{R_{orb}}\la 0.25$~AU) `hot Jupiter' systems jovian-like in all respects apart from the orbital distance would exhibit no main auroral oval auroras.  On the other hand, similar magnetospheres larger than 85~\ensuremath{\mathrm{R_J}} exhibit common current profiles, extending to further distances and thus modestly further toward the pole, depending on the size of the magnetosphere.  The monotonically-decreasing field strength with increasing equatorial radius results in an associated increasing of the current magnitudes toward the pole, and thus precipitating electron energy flux, toward the outer boundary of the model. Increasing the size of the magnetosphere thus increases the emitted radio power, as shown in Fig.~\ref{fig:is}d.\\

\begin{figure}
\noindent\includegraphics[width=84mm]{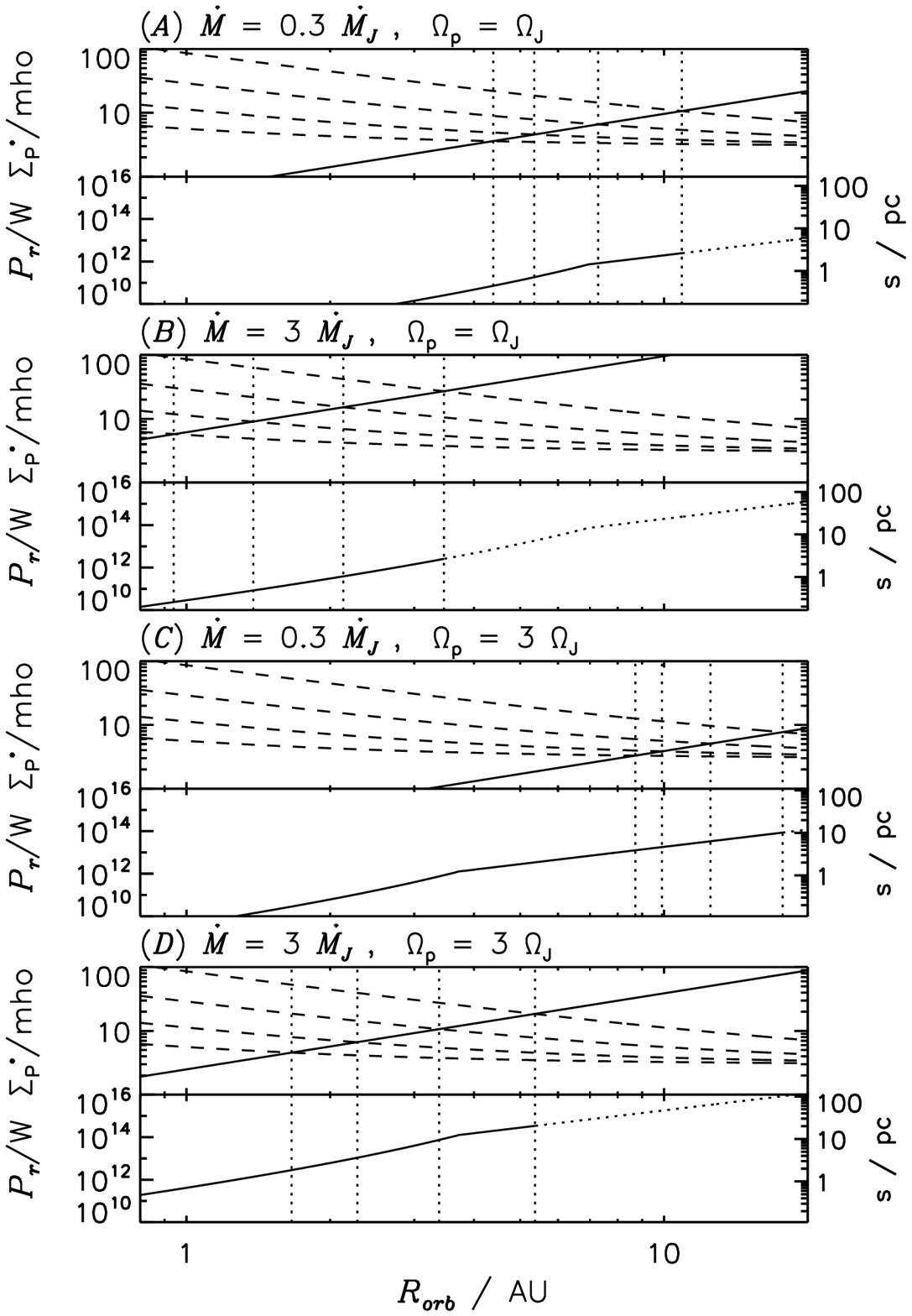}
\caption{
Plot showing results using four combinations of the input parameters \ensuremath{\Omega_p} and \ensuremath{\dot{M}},i.e.\  (a) 
($(\ensuremath{\dot{M}}/\ensuremath{\dot{M}_J})=0.3$, $(\ensuremath{\Omega_p}/\ensuremath{\Omega_J})=1$), (b) 
($(\ensuremath{\dot{M}}/\ensuremath{\dot{M}_J})=3$, $(\ensuremath{\Omega_p}/\ensuremath{\Omega_J})=1$), (c) 
($(\ensuremath{\dot{M}}/\ensuremath{\dot{M}_J})=0.3$, $(\ensuremath{\Omega_p}/\ensuremath{\Omega_J})=3$), and (d) 
($(\ensuremath{\dot{M}}/\ensuremath{\dot{M}_J})=3$, $(\ensuremath{\Omega_p}/\ensuremath{\Omega_J})=3$).  Each section consists of two panels. The top shows the limiting conductance given by Eq.~\ref{eq:spslim} (solid line) and the estimated Pedersen conductance in mho given by Eq.~\ref{eq:sps} for four values of the stellar XUV luminosity, i.e.\  $(L_{XUV\:\star}/L_{XUV\:\sun}) $~=~1,10,100, and 1000  (dashed lines from bottom to top, respectively).  The bottom shows with the left axis the radio power in W emitted, obtained by integrating the precipitating electron energy fluxes using Eqs.~\ref{eq:pe} and \ref{eq:pr}, and with the right axis the maximum distance in parsecs at which sources emitting such powers would be detectable, assuming a detection threshold of 1~mJy.  All parameters are plotted versus orbital distance \ensuremath{R_{orb}}.  The vertical dotted lines in each section indicate where the estimated Pedersen conductances equal the limiting conductance, i.e.\ show the maximum distances \ensuremath{R_{orb}^*} to which the high-conductance approximation is valid, and thus the maximum radio powers for these parameters.  The lines in the bottom of each panel are shown dotted beyond the maximum values of \ensuremath{R_{orb}^*} for $(L_{XUV\:\star}/L_{XUV\:\sun}) = 1000 $, indicating that this region corresponds to XUV luminosities above the upper limit for a young solar-type star as indicated by Ribas et al. (2005) .   
}
\label{fig:sigpow}
\end{figure}
\nocite{ribas05a}
Having briefly examined the effect of magnetospheric size, we now extend the the analysis by considering the parameters \ensuremath{\dot{M}} and \ensuremath{\Omega_p}.  There are presently no published theoretical limitations on the maximum plasma mass outflow rate from exoplanetary moons, which presumably would vary with the number of volcanic moons in a given system as well as the orbital and geological parameters of each satellite and planet.  The theory considered in this paper is valid for all non-zero values of the plasma mass outflow rate, and it is worth noting that at Jupiter the ionosphere and solar wind also act as plasma sources with rates of $<10^2$ and $\ga20$~\ensuremath{\mathrm{kg\;s^{-1}}}, respectively \citep{hill83}, such that in practice the mass outflow rate will probably be non-zero in most systems, even those devoid of active moons.  It therefore seems reasonable to compute the radio power emitted for a range of \ensuremath{\dot{M}} values within an order of magnitude of the canonical jovian figure of 1000~\ensuremath{\mathrm{kg\;s^{-1}}}.  The rotation rate of a planet is governed by its initial angular velocity and subsequent tidal dissipation.  The origin of the former is poorly understood, but candidates include the relative motion of planetesimals during protoplanetary accretion and, for the gas giants specifically, hydrodynamic flows during accretion of hydrogen and helium \citep{lissauer93a}.  A crude estimation of the maximum angular velocity allowed by centrifugal stability of a planet is

\begin{equation}
	\Omega_{max} =\left(\frac{GM_p}{R_p^3}\right)^{1/2}\;\;,
	\label{eq:}
\end{equation}

\noindent where $G$ is the gravitational constant equal to $\ensuremath{\sim}6.67\times10^{-11}\;\mathrm{N\;m^2\;kg^{-2}}$, and which for a planet of Jupiter's mass and radius yields $\Omega_{max} \ensuremath{\simeq}3.3~\ensuremath{\Omega_J}$.  However, Jupiter lies roughly on the boundary between the regime of bodies supported by Coulomb pressure, for which $M_p\propto R_p^3$, and degeneracy pressure, for which  $M_p\propto R_p^{-3}$, such that Jupiter's radius is near maximal for bodies of solar composition. The net effect is that all bodies from solar composition giant planets through brown dwarfs to the very lowest mass stars are expected to have radii similar to Jupiter, and it is thus possible for planets with mass of, e.g.\ $10M_J$ to have angular velocities up to \ensuremath{\sim}11~\ensuremath{\Omega_J}.  The time scale $\tau_{syn}$ required to de-spin a planet by tidal dissipation is given by 

\begin{equation}
	\tau_{syn}\simeq Q\left(\frac{R_p^3}{GM_p}\right)(\Omega_p-\Omega_{orb})
	\left(\frac{M_p}{M_\star}\right)^2 \left(\frac{a}{R_p}\right)^6
	\label{eq:tausyn}
\end{equation}

\noindent where $Q$ is the planet's tidal dissipation factor typically given by $Q=10^5$ for jovian planets,  $\Omega_{orb}$ is the Keplerian angular velocity of the planet, and where we note that factors of order unity have been omitted \citep{goldreich66a,showman02a}. For planets of jovian parameters orbiting solar-type stars at 0.1, 1, and 100~AU, this expression yields tidal synchronisation time scales of $10^8$, $10^{14}$, and $10^{20}$~y, respectively, while for planets more susceptible to tidal locking e.g.\ those with $M_p=0.1M_J$ and $M_{\star}=10M_{\sun}$, the synchronisation time scales are modified to $10^5$, $10^{11}$, and $10^{23}$~y.  Thus, close-orbiting `hot Jupiters' are expected to rapidly become tidally locked, but for planets orbiting beyond 1~AU, tidal effects will be negligible over time scales of e.g.\ the present age of the solar system of $\ensuremath{\sim}4.6\times 10^9$~y, such that the planetary angular velocity may not be much reduced from its initial value at formation.\\

With the above considerations in mind, we now examine the system using four spot combinations of the parameters \ensuremath{\dot{M}} and \ensuremath{\Omega_J}, i.e.\ ($(\ensuremath{\dot{M}}/\ensuremath{\dot{M}_J})=0.3$, $(\ensuremath{\Omega_p}/\ensuremath{\Omega_J})=1$), 
($(\ensuremath{\dot{M}}/\ensuremath{\dot{M}_J})=3$,  $(\ensuremath{\Omega_p}/\ensuremath{\Omega_J})=1$), 
($(\ensuremath{\dot{M}}/\ensuremath{\dot{M}_J})=0.3$, $(\ensuremath{\Omega_p}/\ensuremath{\Omega_J})=3$), and 
($(\ensuremath{\dot{M}}/\ensuremath{\dot{M}_J})=3$, $(\ensuremath{\Omega_p}/\ensuremath{\Omega_J})=3$).  Results are shown in Fig.~\ref{fig:sigpow}, which is divided into four sections of two panels corresponding to the above pairs of parameters.  The solid line in the top panel of each section shows the limiting conductance given by Eq.~\ref{eq:spslim}.  The dependence of the limiting conductance on the parameters \ensuremath{R_{orb}}, \ensuremath{\dot{M}}, and \ensuremath{\Omega_p}, obtained from Eqs.~\ref{eq:bze}, \ref{eq:fe}, \ref{eq:psw}, \ref{eq:hudmod}, and \ref{eq:spslim}, is approximately $\ensuremath{\Sigma_{P\:lim}^*}\propto \ensuremath{\dot{M}}{\ensuremath{\Omega_p}}^{-0.82}{\ensuremath{R_{orb}}}^{1.2}$, such that it increases slightly faster than linearly with distance in each panel, and compared to the values shown Fig.~\ref{fig:sigpow}a, those in Figs.~\ref{fig:sigpow}b, c, and d are thus multiplied by factors of 10, \ensuremath{\sim}0.41, and \ensuremath{\sim}4.1, respectively. The estimated Pedersen conductance is also shown by the dashed lines in each panel for stellar XUV luminosities of $(L_{XUV\:\star}/L_{XUV\:\sun}) $~=~1,10,100, and 1000  (from bottom to top, respectively).  Despite these XUV luminosities being orders of magnitude larger than the solar value, since \ensuremath{\Sigma_P^* \propto (L_{XUV\:\star}/L_{XUV\:\sun})^{1/2}} the conductance values are multiplied from the solar values by factors of only 1, 3.2, 10, and 32, respectively.  However, it is obvious that the increased conductance generated by the higher XUV luminosities greatly increases the maximum orbital distance \ensuremath{R_{orb}^*} at which the high-conductance approximation holds, as shown in each section by the four vertical dotted lines, for example in section (a) at \ensuremath{\sim}4.4, 5.3, 7.3, and 10.9~AU corresponding to conductances of \ensuremath{\sim}3.6, 4.6, 6.6, and 10.6~mho, respectively.  As shown in Figs.~\ref{fig:sigpow}b, c, and d, these values of \ensuremath{R_{orb}^*} are then modified from, e.g., the high XUV luminosity case of \ensuremath{\sim}10.9~AU to \ensuremath{\sim}3.5, 17.7, and 5.4~AU, due to increased \ensuremath{\dot{M}}, \ensuremath{\Omega_p}, and both \ensuremath{\dot{M}} and \ensuremath{\Omega_p}, respectively. \\

The effects of these changes on the maximum radio power are shown in the bottom panels of each section of Fig.~\ref{fig:sigpow}.  The left hand axes indicate the emitted powers, obtained by integrating the precipitating electron energy fluxes using Eqs.~\ref{eq:pe} and \ref{eq:pr}, whilst the right hand axes show the distance $s$ in parsecs obtained from Eqs.~\ref{eq:fr} and \ref{eq:delnu} with these powers and an assumed spectral flux density of 1~mJy.  Note that the right hand axes in Figs.~\ref{fig:sigpow}c and d are different to those in Figs.~\ref{fig:sigpow}a and b, since the spectral flux density is dependent on $\Delta \nu$, which is proportional to  $\ensuremath{\Omega_p}^{3/4}$ through $B_{eq}$. As also shown in Fig.~\ref{fig:is}, the radio power increases with orbital distance, with a slight discontinuity in the gradient where \ensuremath{R_{mp}=85~\ensuremath{\mathrm{R_p}}}.  This discontinuity is simply an artefact introduced by the change in the behaviour of the equatorial magnetic field model at the reference boundary of the `baseline' magnetic field model.  For \ensuremath{R_{mp}<85~\ensuremath{\mathrm{R_p}}}, as the radius of the magnetosphere increases with increasing orbital distance \ensuremath{R_{orb}}, the equatorial magnetic field strength decreases due to flux conservation, such that the current intensities and power output increase as discussed above (c.f.\ the blue and green lines in Fig.~\ref{fig:is}).  On the other hand, as the radius of the magnetosphere further increases such that \ensuremath{R_{mp}>85~\ensuremath{\mathrm{R_p}}} the current intensity at a given co-latitude remains the same, but extends increasingly toward the pole (c.f.\ the red and black lines in Fig.~\ref{fig:is}).  The latter has less of an effect that the former, such that the power output increases less quickly with \ensuremath{R_{orb}} after \ensuremath{R_{mp}} passes 85 \ensuremath{\mathrm{R_p}}.  Thus, for the values of \ensuremath{\dot{M}} and \ensuremath{\Omega_p} in Fig.~\ref{fig:sigpow}a, the maximum radio powers $P_r$ for the four XUV luminosity cases are $\ensuremath{\sim}7.2\times 10^{10}$, $1.8\times 10^{11}$, $8.2\times 10^{11}$, and $2.5\times 10^{12}$~W, corresponding to maximum observable distances $s$ of \ensuremath{\sim}0.4, 0.7, 1.5, and 2.6~pc, respectively.  However, again taking the high XUV luminosity case as an example, this power is modified as shown in Figs.~\ref{fig:sigpow}b, c, and d to $\ensuremath{\sim}2.6\times 10^{12}$, $8.6\times 10^{13}$, and $3.4\times 10^{14}$~W, corresponding to maximum observable distances of \ensuremath{\sim}2.7, 10, and 20~pc, respectively.  Thus, from these four spot combinations of \ensuremath{\dot{M}} and \ensuremath{\Omega_p} it is apparent that the maximum radio power $P_r$ is relatively insensitive to changes in \ensuremath{\dot{M}}, since the higher powers available at a given orbital distance \ensuremath{R_{orb}} for increased \ensuremath{\dot{M}} are compensated for by a decrease in the maximum distance to which the high-conductance approximation is valid \ensuremath{R_{orb}^*}.  On the other hand, whilst changing \ensuremath{\Omega_p} does have competing effects on different components of the system, the overall behaviour is one in which increased \ensuremath{\Omega_p} results in increased power.  We further note that, although, as discussed above, the compressed vacuum dipole relation for the sub-solar magnetopause standoff distance is not strictly appropriate for the rotationally-driven, plasma-filled magnetospheres we consider here, the \cite{huddleston98} relation was derived using observations of jovian values over the range \ensuremath{\sim}45-100~\ensuremath{\mathrm{R_J}}, and extrapolation beyond this range may lead to uncertainties. We have thus also computed the above values using Eq.~\ref{eq:rmpdip} in place of Eq.~\ref{eq:hudmod}, and find that the results are qualitatively similar, except that the values of \ensuremath{R_{orb}^*}, $P_r$, and $s$ are modified by factors of \ensuremath{\sim}2-4, \ensuremath{\sim}0.1-0.4, and \ensuremath{\sim}0.3-0.6, respectively.  As noted by \cite{cowley02} and \cite{nichols03}, the effect of the stretching of the planetary field from a dipolar configuration into a magnetodisc structure is to amplify the upward field-aligned current density associated with the aurora and radio emissions by 1-2 orders of magnitude, such that the precipitating electron energy flux, a quantity proportional to the square of the field-aligned current density through Eq.~\ref{eq:ef}, is modified by 2-4 orders of magnitude.  Using Eq.~\ref{eq:rmpdip} in place of Eq.~\ref{eq:hudmod} compresses the equatorial magnetic field to a more dipole-like form, such that the radio power available at a given orbital radius is significantly reduced.  However, the effect of a closer boundary is also to reduce the conductance required to maintain near-rigid corotation throughout the magnetosphere, such that the maximum power reduction is partly mitigated by increased orbital radius at which the maximum powers are available.  However, we again note that Eq.~\ref{eq:rmpdip} does not reproduce results consistent with observations of Jupiter's magnetosphere, such that in the discussion that follows we employ results obtained using the \cite{huddleston98} relation.  \\

\begin{figure}
\noindent\includegraphics[width=84mm]{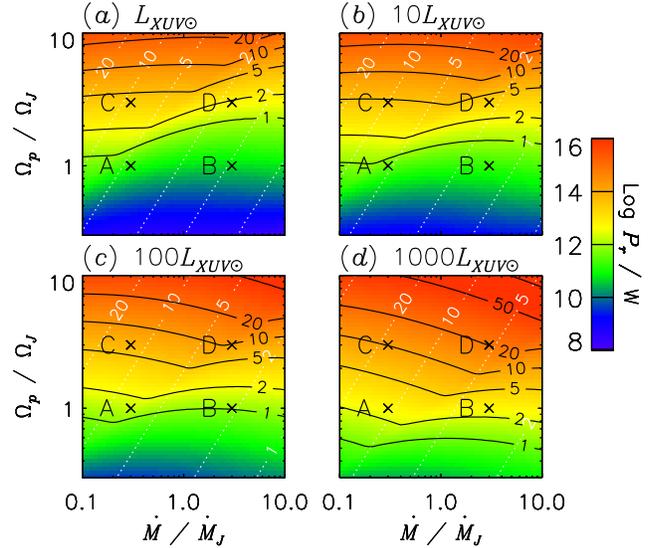}
\caption{
Coloured plots indicating the maximum radio powers $P_r$ available using different pairs of system parameters \ensuremath{\dot{M}} and \ensuremath{\Omega_p} for, in panels (a)-(d) respectively, the XUV luminosity cases $(L_{XUV\:\star}/L_{XUV\:\sun}) $~=~1,10,100, and 1000.  Also shown by the black contours are the maximum distances $s$ in pc at which sources of these powers are observable, assuming a detection threshold of 1~mJy, and the white contours show the orbital distances \ensuremath{R_{orb}^*} in AU at which these maximum powers are available.  Labelled crosses refer to the spot combinations shown in Fig.~\ref{fig:sigpow}.
}
\label{fig:grids}
\end{figure} 
 
We now extend the analysis to cover the full parameter space within the considerations discussed above. We thus examine the behaviour of the system over $0.1<(\ensuremath{\dot{M}}/\ensuremath{\dot{M}_J})<10$ and $0.3<(\ensuremath{\Omega_p}/\ensuremath{\Omega_J})<10$.  We should note that angular velocities above \ensuremath{\sim}3~\ensuremath{\Omega_J} would require planets more massive than Jupiter and may thus have larger dipole moments, although as discussed above for simplicity we do not take this into account here and will be the subject of future study.  Results are shown in Fig.~\ref{fig:grids} for the four different XUV luminosity cases.  The colour indicates the maximum radio power $P_r$ available for each pair of parameter values, as shown in Fig.~\ref{fig:sigpow} by the powers at the locations of the vertical dotted lines, and the black contours show the maximum observable distances of sources emitting such radio powers, again assuming a spectral flux density threshold of 1~mJy.  Note the kinks are an artefact of the model indicating the change of behaviour where \ensuremath{R_{mp}=85~\ensuremath{\mathrm{R_J}}} as discussed above.  The white contours indicate the orbital distances \ensuremath{R_{orb}^*} at which these powers are available, which we note increase with decreasing \ensuremath{\dot{M}} and, to a lesser extent with increasing \ensuremath{\Omega_p}, as previously noted.  This plot confirms the overall behaviour apparent in Fig.~\ref{fig:sigpow}, i.e.\ that the radio power is essentially independent of \ensuremath{\dot{M}}, increases with \ensuremath{\Omega_J}, and increases with stellar XUV luminosity.  Thus, for stars with solar XUV luminosity, planets with $(\ensuremath{\Omega_p}/\ensuremath{\Omega_J})\sim5$ are required to produce radio emissions detectable from beyond \ensuremath{\sim}10~pc, but this is reduced to e.g.\	$(\ensuremath{\Omega_p}/\ensuremath{\Omega_J})\simeq 2$ for stars with $(L_{XUV\:\star}/L_{XUV\:\sun}) $~=~1000 and $(\ensuremath{\dot{M}}/\ensuremath{\dot{M}_J})\simeq 2$.  In all XUV luminosity cases, a significant number of parameter combinations within an order of magnitude of the jovian values are capable of producing emissions observable beyond 1~pc, in most cases requiring exoplanets orbiting at distances between \ensuremath{\sim}1 and 50~AU. For the two higher XUV luminosity cases, parameter combinations within an order of magnitude of Jupiter's could generate emissions detectable beyond \ensuremath{\sim}50~pc.

\section{Summary and discussion}

In this paper we have provided the first consideration of magnetosphere-ionosphere coupling at Jupiter-like exoplanets.  We have estimated the radio power emitted by such systems under the condition of near-rigid corotation throughout (a condition which maximises the field-aligned currents and thus the radio power for a given magnetosphere), in order to examine the behaviour of the best candidates for detecting internally-generated radio emission with next generation radio telescopes such as LOFAR.  We have thus estimated for different stellar XUV luminosity cases the orbital distances within which the ionospheric Pedersen conductance would be high enough to maintain near-rigid corotation, and we have then considered the magnitudes of the large-scale magnetosphere-ionosphere currents flowing within the systems, and the resulting radio powers, at such distances.  We have also examined the effects of two key system parameters, i.e.\ the planetary angular velocity and the plasma mass outflow rate.\\

The key results of the study can be summarised as follows:

\begin{enumerate}
	\item The radio power emitted increases with increasing system size, and thus increases with orbital distance within the limit of validity of the high-conductance approximation.
	\item The limiting orbital distance, which defines the maximum radio power available for a given set of system parameters, increases with stellar XUV luminosity and planetary rotation rate, and decreases with magnetospheric plasma mass outflow rate.
	\item The overall effect is that the radio power emitted increases with planetary rotation rate, but is essentially independent of plasma mass outflow rate since the higher powers available at a given orbital distance for increased plasma mass outflow rate are compensated for by a decrease in the maximum orbital distance to which the high-conductance approximation is valid.
	\item In all XUV luminosity cases studied, a significant number of parameter combinations within an order of magnitude of the jovian values are capable of producing emissions observable beyond 1~pc, in most cases requiring exoplanets orbiting at distances between \ensuremath{\sim}1 and 50~AU. For the higher XUV luminosity cases the observable distances for jovian mass planets can reach \ensuremath{\sim}20~pc, and massive, rapidly rotating planets could be detectable beyond \ensuremath{\sim}50~pc. 
\end{enumerate}

However, we should note here the limitations of the simple model used in this study.  First, we have not considered how the structure of the magnetic field changes with planetary angular velocity and plasma mass outflow rate.  As noted above, Jupiter's magnetosphere is partly inflated by the centrifugal force of iogenic plasma \citep{caudal86}, such that systems with higher angular velocity or plasma loading would be expected to be further inflated, although this would be somewhat mitigated by the associated increase of the planetary field strength in the case of the angular velocity.  However, as discussed in Section~\ref{sec:results}, the effect of the stretching of the planetary field from a dipolar configuration into a magnetodisc structure is to amplify the upward field-aligned current density associated with the aurora and radio emissions by 1-2 orders of magnitude, and therefore the power values derived here may be viewed as lower limits in this regard. Second, we note that the field-aligned voltages obtained by the current-voltage relation used in this study may underestimate the true voltages if the location of the accelerator is fixed at a few planetary radii up the the field lines, such that these power values might again be underestimates.  In a related point, we note that simply defining the plasma angular velocity as we have done ignores the effects of the significant field-aligned voltages, which act to de-couple the equatorial and ionospheric plasma angular velocities \citep{nichols05, ray10a}, an issue which should be studied in further works.  Further, we note that we have not considered at all the stellar wind interaction mediated by magnetic reconnection at the dayside magnetopause, which at Jupiter could be associated with the many variable and sometimes extremely bright polar auroras observed \citep{waite01,pallier01, grodent03a, bunce04, nichols09a,nichols09b}, and would thus sporadically increase the power output. Nor have we considered the effect of changing stellar wind dynamic pressure, which at Jupiter is known to modulate the intensity of the UV and radio emissions by factors of \ensuremath{\sim}3 \citep{gurnett02,cowley07,clarke09a,nichols09b}, and again, although the details of how this affects Jupiter's auroral and radio emissions remain to be fully determined, this effect may also act to increase the output powers from those computed here.  However, we also reiterate that our assumption of solar values for the stellar wind velocity and density may represent underestimates for stars more active than the Sun, such as those much younger.  Higher stellar wind dynamic pressure values would decrease the size of the magnetosphere for a given orbital distance, thus reducing the emitted radio power. It is clear that the evolution of internally-generated radio emissions over the lifetime of a star should therefore be considered in future studies.  We should also recall that we have only considered here planetary magnetic fields with the same polarity as those of Jupiter and Saturn, i.e.\ with magnetic and spin axes co-aligned to first order, since this is the configuration that has been most studied for bodies in our solar system.   However, it is probable that only 50\% of Jupiter-like exoplanets exhibit this polarity.  The effect of reversed polarity is to reverse the direction of the current system, such that downward currents are replaced with upward currents and vice versa, such that the source population for the field-aligned currents in this case may be very different to that considered here (see e.g.\ \cite{bunce04} for a discussion of field-aligned currents induced at Jupiter's dayside magnetopause).  The emissions from planets with the opposite polarity should be studied in future works. \\

Finally, we note that the implication of these results is that the best candidates for detection of such internally-generated radio emissions are rapidly rotating Jupiter-like exoplanets orbiting stars with high XUV luminosity at orbital distances beyond \ensuremath{\sim}1~AU.  This type of exoplanet has not previously been considered as potential detection candidates for next generation radio telescopes, but searching for such emissions such may offer a new method of detection of more distant-orbiting exoplanets less likely to be detected by those techniques which are biased toward close-orbiting `hot Jupiters'.  However, dual detections with radio telescopes and conventional methods would best constrain the planetary parameters.

\vspace{-0.5cm}

\section*{Acknowledgments}

JDN was supported by STFC Grant ST/H002480/1, and wishes to thank S.~W.~H. Cowley, M.~A. Barstow, M.~R. Burleigh and G.~A. Wynn for constructive discussions during this study, and also thanks the referee for providing helpful comments on the manuscript.

% \bibliographystyle{mn2efixed}
% \bibliography{../references}

\newpage

\label{lastpage}

\end{document}